# A Skyline and ranking query odyssey: a journey from skyline and ranking queries up to f-skyline queries

Giuseppe Sorrentino

Politecnico di Milano
Milan, Italy
giuseppe.sorrentino@mail.polimi.it

**Abstract**

Skyline and ranking queries are two of the most used tools to manage large data sets. The former is based on non-dominance, while the latter on a scoring function. Despite their effectiveness, they have some drawbacks like the result size or the need for a utility function that must be taken into account. To do this, in the last years, new kinds of queries, called flexible skyline queries, have been developed. In the present article, a description of skyline and ranking queries, f-skyline queries and a comparison among them are provided to highlight the improvements achieved and how some limitations have been overcome.

## 1 Introduction

Nowadays, being able to handle large data sets is essential for several different applications. Two of the most used queries to retrieve data from databases are skyline queries and ranking queries (or top-k queries). The first one is used to retrieve tuples that are non-dominated in a specific dataset. Later in the article, the meaning of domination is better described. Their name evokes their shape since the output looks like a skyline. Every point in the skyline represents the best element according to some criteria.

The second kind of query allows the retrieval of the top-k records in a dataset. Since the result of this tool is a ranking of tuples, it has several applications, e.g. votes. In a poll where each voter expresses votes or preferences on candidates, the top-k query is employed to retrieve the result ranking of the election.

Due to the frequent use of this query, avoiding as many downsides as possible is essential. Actually, despite the effectiveness of these queries, their disadvantages have to be taken into consideration. For example, skyline queries could return a huge size result (which is not admissible in most cases), while top-k ones require a utility function, which is far from simple to specify. Since the importance of skyline and ranking queries is undeniable, different versions of these tools have been developed and this article is structured to lay the groundwork for comparison among consolidated and upgraded versions of them.

In the first section, there is a characterization of different kinds of skyline and ranking queries. After the description, their strengths and weaknesses are stressed. Then, there is the description of variants, called flexible skyline queries[4], whose strengths and weaknesses are analyzed too.

Based on these two descriptive sections, in the last part of the paper, there is a comparison between these two kinds of queries aimed to highlight the improvements achieved thanks to flexible skyline queries. Precisely, this comparison is the real goal of the article since it allows us to prove the importance of the improvement reached and their effectiveness.

# 2  Skyline Query

One of the most challenging problems concerns the determination of preferences from a specific dataset, which is provided with several attributes. This is also called the "multi-criteria decision making" problem. There are two different ways to face this, and they are grounded on various ideas.

On the one hand, there's a first paradigm considering domination. A record of a dataset dominates the others if it represents the optimal option [4] according to some criteria. So, considering the attributes of the records and a preference criterion, a record is said to dominate another one if it better fits those criteria. For the sake of clarity, let's consider these two records in table 1:

| Name | Role | Age | goal scored last year |
|------|------|-----|----------------------|
| A | Midfielder | 18 | 100 |
| B | Striker | 18 | 80 |

Table 1: Example of records

If the scout wanted a player that scored a lot last year, the record A dominates B since the attribute "goal scored last year" has a higher value.

Based on this concept, skyline queries are employed to retrieve records that are not dominated by any other record in the dataset. Since they can be too strict for common applications, several alternatives like the k-skyband queries, which include all the elements dominated at most by k-1 elements, are developed.

Although the advantage of this tool is evident, since they are intuitive and they retrieve several precise information, the downsides are not insignificant. Indeed, output size could be huge and cannot be customized. On the other hand, there is the possibility of considering ranking instead of domination. This method allows evaluating each record with a score computed by a specific user function. Thanks to this, preferred records can be obtained using a top-k query. Notwithstanding, this method is far from easy since specifying the correct user function for scores might be hard [8] but further details are provided in the next section. In the following paragraphs, there is a description of the main kinds of the skyline and top-k queries, while, later on, the focus moves to the results achieved with f-skyline queries.

## 2.1  Progressive Skyline

Although this tool raises huge interest among the scientific and database community, the algorithms developed contain some serious assumptions and constraints which limit their applicability.

One of the most stressed methods in practical applications is the progressive one, which can retrieve a result without reading the entire database. This approach is preferred since it allows to work with enormous databases avoiding a too high computational cost. In [7] is shown branch and bound skyline, also called BBS, based on nearest-neighbour search, which is a form of proximity search used in the optimization problem of finding the point, in a given set, that is closest (or most similar) to a certain specific point. Closeness is typically expressed in terms of a discordant function: the less similar the objects, the larger the function values.

The robustness of this approach is the I/O optimality while deficiencies consist of the inability of reducing the size of the result and the impossibility of customizing the query with user preferences. A detailed example of this method is also described in [8]. To clarify it, let's state the main strengths of the algorithm: the possibility of computing and upgrading a skyline query progressively. The importance of this is shown in the following example considering a possible upgrade on a dataset.

In table 2 a dataset of football players is provided while in figure 1 its skyline is plotted:

| Name | Yellow cards in the last year | red cards in the last year |
|------|-------------------------------|----------------------------|
| A    | 1                             | 6                          |
| B    | 3                             | 4                          |
| C    | 5                             | 3                          |
| D    | 7                             | 1                          |
| E    | 4                             | 3                          |

Table 2: Dataset used for computing progressive skyline query

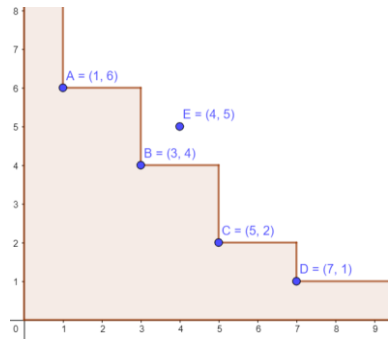

Figure 1: Output of the skyline query

Now, let's consider the insertion of a new tuple P in the dataset having 2 yellow cards and 3 red cards in the last year.
The first operation performed by the BBS is a dominance test on the dataset. If the record is dominated by the already present skyline, it is simply discarded. If the record is not dominated, then it represents a new point of the skyline query. After this check, the BBS algorithm has to control which records are dominated by the new one. If no records are dominated, the size of the result simply increases by one. If there are previously non dominated elements which are now dominated by the new one, they must be removed from the skyline.
This second case is shown in Figure 2, where on the left the previously computed skyline query is shown with the new element P added in the non-dominated area while on the right there is the new skyline query where the record B has been removed (since it is dominated by P).

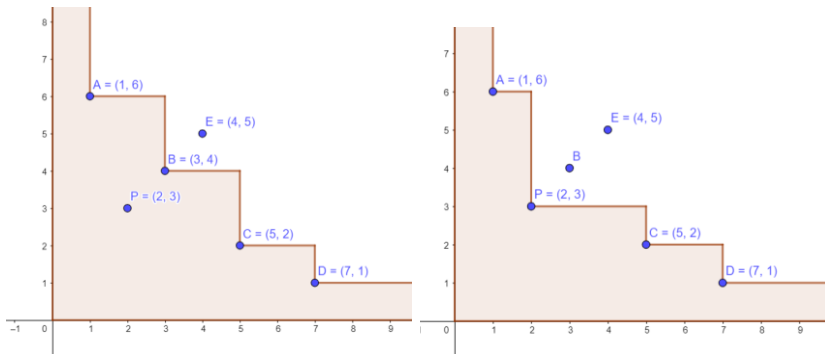

Figure 2: (left) non updated Skyline query with P, (right) updated skyline query with P added

Just for completeness, another possible update must be considered: the deletion of a record from the dataset. If the record deleted is a dominated one, nothing changes, but several considerations are needed if the deleted one is a point of the skyline. In fact, if the eliminated record is non dominated, all the previously dominated points may now be non dominated and they could be in the skyline. To show this, in Figure 3 the first presented skyline query is now shown without the record B which has been deleted. Thanks to the deletion, the record E is now part of the skyline output

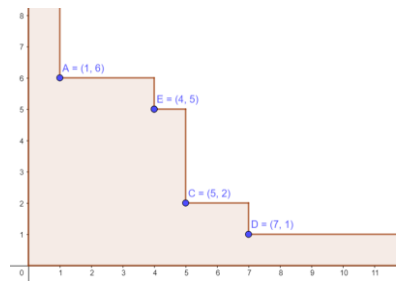

Figure 3: Output of the skyline query after the deletion of B

Since the deletion may be a computationally expensive operation, several optimization techniques are possible like the use of a buffer to store deleted tuples, in case of their future reuse, or the organization of data in an R-tree [7].

## 2.2 P-Skyline

In several contexts focused on data, including data mining and database systems, there are three possible approaches to deal with multi-objective optimization. Two of them have been previously presented, the skyline query and ranking query previously mentioned, the last one is called lexicographical approach. It consists in establishing a strict priority among the attributes. As previously argued, none of these approaches is completely suitable, so others are developed trying to achieve a better result.

Aimed to reduce the output size, in [17] the authors introduce the prioritized skyline query. These queries are thought to merge the benefits of skyline queries and the lexicographical method.

The basic idea of the lexicographical approach is that considering two attributes "A" and "B": if "A" is more important than "B" then each tuple with a better value of "A" is unconditionally preferred to each tuple with a worse value of the same attribute. According to this, p-skyline and lexicographic approach may seem like the same tool, but there is an important difference: in p-skyline there is the possibility of considering also different kinds of criteria. For this reason, this new kind of skyline can solve each problem solvable with the lexicographic approach and even different problems to which this approach can't be applied.

Considering this, it must be specified that, in the case of equally important attributes, this query is equivalent to the classic skyline framework. Analysing this method, the main benefit results to be the reduction of the output size while the main downside is the impossibility of a trade-off. Actually, p-skylines can be seen as a special case of non-compensatory decision making, thus there can't be any compensation among attributes and the user cannot specify any trade-off[19].

## 2.3 Continuous skylines

If a query involves not only static dimensions but also dynamic ones it is called continuous skyline. To deal with this query, a computation over streaming datasets is done to produce a skyline summary over time [8]. While common skylines work on static data sets, the continuous time-interval skyline operation involves data points valid for a specific time interval. For this reason, they have an arrival time and an expiration time. The task of the DBMS is to compute this skyline continuously for each data point that is valid at a given time. There are a lot of common situations in which this kind of query is useful. For example, also for hotels or voyages, there are often some prices valid only for specific periods.

| Name | Starting Time | Expiration Time | price |
|---|---|---|---|
| Hotel 1 | 01-05-2022 | 31-05-2022 | e25 per day |
| Hotel 2 | 07-05-2022 | 22-05-2022 | e20 per day |

Table 3: Example of records of a dataset which are valid only for a specific time-period

The first algorithm for efficiently evaluating the continuous time-interval skyline operation is called LookOut and it is described in [14]. Considering the dataset in table 4 and the two records mentioned above:

| Name | price |
|---|---|
| Hotel 1 | e140 per day |
| Hotel 2 | e145 per day |
| Hotel 3 | e40 per day |
| Hotel 4 | e45 per day |

Table 4: Dataset about hotels without special prizes

As it can be noticed, the hotel3 is not dominated by any other record but, considering the special prizes presented in Table 3, the algorithm returns also the hotel 2 for the interval from 07-05-2022 up to 22-05-2022 and hotel 1 for the other days in which the special prize of hotel 1 is valid. The strengths of the LookOut algorithm are the time and space efficiency beyond great scalability but it does not enable the continuous skyline query to overcome all the weaknesses of common skylines like the huge size of the result.

## 2.4 Subspace skyline

Another relevant problem concerns employing skyline queries in a subspace of the data. To understand the meaning of subspace let's consider the following points:
A = (4;1)
B = (3;3)
C = (1;5)
Since these data are described in 2 dimensions, there are 2 different subspaces, the subspace X and the subspace Y. Considering this, an n-dimensional space has several subspaces, each one is described by a partial view of the entire space[10]. In the above-mentioned article, an efficient algorithm, SKYEY, is presented to show how to take into account the existence of these subspaces and which advantages are obtained by them. Using this solution, which employs a top-down approach, the skyline in the subspace is recursively computed. Due to the computational efforts, data have to be managed using some strategies like sorting. Other approaches are the SKYCUBE [24] and the SUBSKY [25]. In particular, the last approach is relevant since it allows transforming a multidimensional data set into a mono-dimensional one. This modification is powerful since these new data can be indexed with a single B-tree, implemented in any relational database. However, there are several troubles with this kind of query [8]. For example, obtaining the meaningful subset of skyline points in any subspace remains a challenging task in the current mobile internet. Moreover, the computational cost of these skyline queries may represent a shortcoming when it provides numerous candidate attributes. Actually, in many applications, various users may focus their attention on different subsets of the attributes according to their interests and computing effort dramatically increases.

# 3 Top-k Queries

As previously mentioned, an alternative to the skyline query is the ranking query, also called the "top-k" query. The objective of this query is to find the best K objects that satisfy the user's need [9]. This needing is formulated as a scoring function over the object's attribute values. The use of this scoring function allows reducing the original multi-objective problem to a single-objective one [3].

Another similar approach to select a limited subset of skyline records is to assign them a measure of importance based on specific properties. The result of this approach is the set of top-k representative skyline points, also called RSP. In [23] its issue is formally defined, but this approach, like skyline queries presented in the previous section, does not allow dealing with the deficiencies discussed in solving the "multi-criteria decision making" problem. Moreover, the task of finding the RSP is NP-Hard and only approximated solutions can be used in practice [3].

For the sake of clarity, the following example shows the effects of top-k queries and the main drawback: how to identify the scoring function. we assume the existence of a scoring function S and its ability to evaluate a football player returning an integer value from 1 to 100. Ignoring how this function works, we consider the dataset in table 5:

| Name | Role | PAC | SHO | PAS | DRI | DEF | PHY |
|------|------|-----|-----|-----|-----|-----|-----|
| AAA | Striker | 50 | 79 | 60 | 40 | 55 | 90 |
| BBB | Striker | 80 | 90 | 90 | 60 | 50 | 80 |
| CCC | Striker | 75 | 75 | 90 | 90 | 60 | 70 |
| DDD | Striker | 22 | 33 | 38 | 80 | 90 | 66 |

Table 5: Dataset of Football Players

Now, the aforementioned function is handled. It evaluates players and returns the two best players above all. In this case, it is evident that the result depends on the function and this function must take into account several elements. Indeed, a goalkeeper should not be considered a worse player than a striker for a smaller number of goals scored. So there must be different criteria. In table 6, the score for each player is shown, computed as the average of each attribute, while the coloured rows represent the output of the top-k query with K = 2.

| Name | Role | Score |
|------|------|-------|
| AAA | Striker | 62 |
| BBB | Striker | 75 |
| CCC | Striker | 76 |
| DDD | Striker | 54 |

Table 6: Result of scoring function and result of the top-k query with K = 2

As can be easily imagined, finding a function like the S function mentioned before is extremely hard. For example, neither the exemplifying S used above is effective since in the average of strikers also parameters like defensive abilities are considered as important as the others, but this is quite senseless. Looking at this example, the shortcomings of ranking queries are relevant and remarkable.

# 4 Flexible skyline

Each one of the previously mentioned queries has its advantages and allows us to solve a particular problem. However, they share some disadvantages which must be solved as shown in [12].
First of all, skyline queries do not permit the use of customer preferences since neither the compensation of more important attributes with less important ones nor the customization of user preference is possible. In addition to them, there are also the top-k queries that take into account user preferences, but require the use of a scoring function that is often uneasy to define.
In [15, 2, 21, 6, 18, 12] different solutions are described, using new kind of skyline queries.

## 4.1 Definitions

The former elements added to face the disadvantages of common skyline queries are new definitions of domination. Several of them have been proposed combined with some specific methods, like the F-dominance for the ND and PO operators in [3], the $\epsilon$ - dominance for the $\epsilon$-skyline in [21] or the $\rho$-dominance for the ORU and ORD operators in [12]. All these definitions differ for some features which characterize the method they are used for. In the following sections, each description is forerun by the proper definition of dominance. As stated in [3], these new methods aim to merge the benefits of skyline queries and top-k queries, to face all the issues underlined in the previous section of this paper. In the following section, some of the most important kinds of flexible skylines are introduced, to lay the groundwork for a comparison performed in the last chapter.

## 4.2 K-regret query

All the main methods known to solve the multi-criteria decision problem have some downsides due to the high size of the result ( for the skyline queries) or the need for a scoring function provided by the user ( for the top-k queries). Such limitations are significant since they constrain a lot the applicability of the methods.

Recently, to maintain top-k queries' features and face its issues, a different kind of query has been introduced in [15]: the k-regret queries.

It is based on returning records which minimizes the maximum regret ratio. As stated in [16], the regret ratio quantifies the amount of disappointment of a user who gets the best record of a specific subset but not the best one possible.

In [15] a lower bound of this regret ratio is shown and justified while the upper bound is described using an efficient algorithm, called SPHERE. The upper bound described in this algorithm is shown to be asymptotically optimal and restriction-free for any dimensionality, so it represents one of the best results in the literature.

The main strength of the K-regret query is the absence of a score function provided by the user, which represents the main inadequacy of each kind of top-k query. Avoiding it, all the main lacks of top-k queries are faced without renouncing their merits.

Nowadays, thanks to its flexibility, this tool has great applicability in many fields like Information Retrieval [20] or Recommendation Systems [11] where they can recommend something according to some criteria without choosing a specific utility function.

In the following lines, a simple example to explain what is the maximum regret ratio and what is the real goal of this tool is provided.

For the sake of clarity, this table shows a dataset about unreal football players and their statistics:

| Name | Role | Assist | Goal |
|------|---------|--------|------|
| AAA  | Striker | 18     | 50   |
| BBB  | Striker | 22     | 33   |
| CCC  | Striker | 44     | 8    |
| DDD  | Striker | 30     | 49   |
| EEE  | Striker | 25     | 44   |
| FFF  | Striker | 18     | 40   |

Table 7: starting dataset

Then, considering a pool of possible utility functions, for each tuple of the dataset is calculated a specific value simply using the following function:

$$F = W_1 * Assist + W_2 * Goal$$

| Name | Role | Age | Assist | Goal | $F_{0.4;0.6}$ | $F_{0.7;0.3}$ | $F_{0.3;0.7}$ |
|------|------|-----|--------|------|---------------|---------------|---------------|
| AAA | Striker | 22 | 18 | 50 | 37.2 | 27.6 | 40.54 |
| BBB | Striker | 25 | 22 | 33 | 28.6 | 25.3 | 29.7 |
| CCC | Striker | 26 | 44 | 8 | 22.54 | 33.2 | 18.8 |
| DDD | Striker | 28 | 30 | 49 | 41.4 | 35.7 | 43.3 |
| EEE | Striker | 18 | 25 | 44 | 36.4 | 30.7 | 38.3 |
| FFF | Striker | 24 | 18 | 40 | 31.2 | 24.6 | 33.54 |

Table 8: Utility functions applied to the dataset

The coloured rows consider the subset K of under 25 players. For each utility function, the regret ratio has been computed, using this formula:

$$\frac{max_{p \in Dataset} - max_{p \in K} f(p)}{max_{p \in Dataset}}$$

So for the K records of the previous table all the scores are presented in table 9:

| Function | Name | Regret Ratio |
|----------|------|--------------|
| $F_{0.4;0.6}$ | AAA | $\frac{41.4 - 37.2}{44.1} = 0.1014$ |
| $F_{0.7;0.3}$ | EEE | $\frac{35.7 - 30.7}{35.7} = 0.1400$ |
| $F_{0.3;0.7}$ | AAA | $\frac{43.3 - 40.54}{43.3} = 0.0637$ |

Table 9: Regret Ratio for each utility function

As a result of the computation, the next step of the algorithm consists of calculating the maximum regret ratio which is the biggest ratio among the previously calculated. In this case, for example, the Maximum regret ratio would be 0.14. In conclusion, the idea of SPHERE and the k-regret queries is retrieving a subset of K tuples which has the minimized maximum regret ratio, to narrow the choice from a small subset that probably will satisfy the user.

## 4.3 Trade-off Skyline

Dealing with the real world means dealing with trade-offs. In the majority of cases, there is no way of doing something without accepting some compromises and the trade-off skylines are exactly based on this idea. Based on the Pareto semantics, an algorithm capable of taking into account trade-offs has been developed [2]. After formalizing the concept of trade-offs, the authors show an algorithm that reduces the size of the results and, at the same time, permits compensation. Moreover, a lot of methods for pruning to obtain a result that is redundant free are presented. In particular, if each attribute of an object A can be defined better or equal to each corresponding attribute of an object B, then it is possible to state that B can be pruned by the result [13].

While common skyline can be tested and optimized with domination, this kind of query needs a different approach in which the domination test is replaced by component-wise comparisons [13]. In order to clarify how this tool works, here a simple example is provided which exposes what a trade-off is and how it can be used. By definition, a skyline query returns a set of tuples that are non-dominated by others. For this reason, these tuples must be considered incomparable and, as previously said, there could be a huge amount of elements as result. Notwithstanding, if there was an additional constraint on the output, there would probably be the possibility of comparison among these tuples and thus a reduction of the output.

This constraint is the trade-off. Looking at the figure 4, it is possible to see how can be "represented" the trade-off.

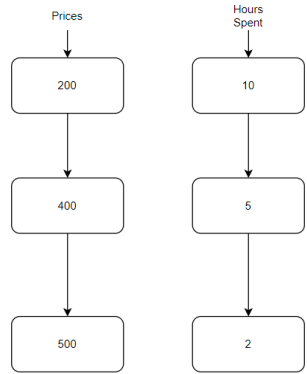

Figure 4: Example of Skyline query

To understand this figure, let's picture a case of a user who wants to book a voyage. He would ask for help from the travel company and it would give them this trade-off: he would be willing to spend 200 euros taking no more than 10 hours to reach the destination. Notwithstanding, he would be satisfied also spending 400 euros, provided that no more than 5 hours are required to reach the destination or he would be fulfilled even spending 500 euros, provided that the travel lasts no more than 2 hours. All this information can be represented as a constraint for the trade-off query which can reduce the size of the output providing a result that is more satisfying for the user and so providing the tuples that dominate the other w.r.t. trade-off set. Considering this example, in figure 4 two elements belonging to the trade-off set are represented and they can be used by the tool in the following way:

1. Apply the Skyline to the dataset
2. Consider the trade-off provided by the user, in this case, the trade-off is represented in Figure 4
3. Cut off the output to fit the constraint represented by the trade-off

Actually, this is not the more efficient way to compute this query. In fact, the best algorithm provided evaluates the trade-off constraint on the fly [13]. Despite this, considering this example it is clarified how this kind of skyline query can face all the main downsides of non-flexible skyline queries without having remarkable disadvantages.

## 4.4 the $\epsilon$ - skyline

As already described in the introductory section, one of the most common strategies to improve skyline queries and face their defects is the definition of a new dominance rule. In this section, a particular kind of solution called $\epsilon$ - skyline query is shown and it is based on a concept of dominance called $\epsilon$ - dominance [21]. This expands the general meaning of domination without differing too much from it. The main idea is the following: Given a dimensionality d, a tuple $t_1$ $\epsilon$ -dominate a tuple $t_2$ if

- given a set of tuples T with T[1]...T[d] attributes,
- given a set of weights W with W[1]...W[d] values,
- given a value $\epsilon$ belonging to [-1,1],

For each i belonging to [1,d] then
$$t_1[i] * W_1[i] <= t_2[i] * W_2[i] + \epsilon$$
Essentially, when considering the domination region a partial area according to the value of epsilon is added. This means that a positive value of $\epsilon$ will lead to a smaller returned dataset while a negative value of $\epsilon$ will lead to a bigger returned dataset. In the following example, the basic concept is presented. The next figure

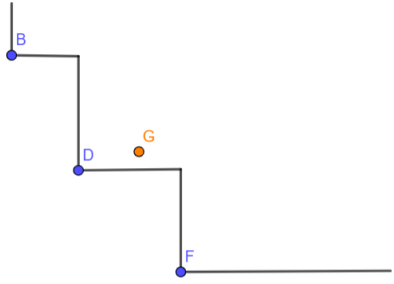

Figure 5: Example of Skyline query

represents a common skyline query:

Conversely, in figure 6, there is the qualitative application of the $\epsilon$-skyline query.

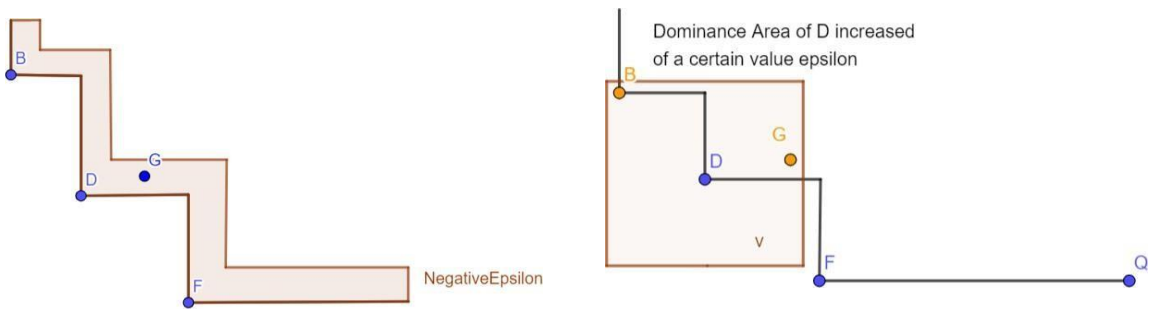

Figure 6: $\epsilon$-skyline query with negative (left) and positive (right) $\epsilon$ for D

On the left, considering a negative epsilon, it is evident how the resulting dataset is increased since the record G is not dominated by any other point. As opposite, on the right, considering a positive epsilon, it is evident that the domination region of D now includes also points B and G so they are both dominated by D thus they are not part of the $\epsilon$-skyline result.

The proper value of $\epsilon$ allows to model the query according to preferences, facing all the main drawbacks of common skyline queries like the huge size of the result. In particular, normalizing each dimension for the sake of simplicity, we can state that by increasing $\epsilon$ from -1 to 1 there is a monotonic increase of the result, which goes from containing 0 tuples to containing the entire dataset [21].

## 4.5 ND and PO operators

As previously stated at the very beginning, the ND and PO operators' basic idea is the definition of a new kind of dominance, the F-Dominance. As formally described in [6], F is a family of different scoring functions and, while a tuple is said to dominate another if all its attributes are more preferable [4], a record X is said to F-dominate another record Y when X is always better than Y according to every function in the family F. To each tuple X is so assigned a dominance region which is defined as the set of points which are F-dominated by X according to the previously defined F, in a d-dimensional space[0,1].

The first operator which is used to show the idea of F-Dominance is the **ND operator**. A formal detailed description is present in [3] but the purpose is to return a set of records that are non-F-dominated by any other record. Many properties of this operator are based on its monotonicity. The ND is a monotone operator with respect to the set of scoring functions and a consequence of this is the growth of the F-dominance region for a smaller set of functions. In [6] the computation of the ND operator is shown. Since it has already been stated that it returns a subset of the skyline, it can be obtained after computing the skyline query with any one of the well-known algorithms. Alternatively, it can directly be computed from the dataset.

The other two relevant properties presented by the authors of the aforementioned article are the entailment of dominance by F-dominance and the transitivity of F-dominance. Their importance relies on the evaluation of different tuples of the dataset since they allow the reduction of the computational cost of the query.

A second operator is the **PO operator**. It is described in [3] and it returns all the potentially optimal records. A record is said to be potentially optimal if it is optimal for at least a function of the family F. Besides, PO properties are formally described in [6] but the main ones, like the monotonicity, are also valid. Similarly to the ND operator, even for the PO several alternatives are available. It can be estimated after the computation of the ND operator or it can be directly obtained from the input data. In particular, since obtaining the potentially optimal tuples from the dataset may be costly in the article also an optimization technique has been introduced.

The effectiveness of these F-skyline queries has been evaluated and the reduction of the size of the result is proved. Thanks to this, one of the main problems of the common query is mostly solved. In particular, the authors demonstrate the effectiveness of these F-Skyline operators by calculating the ratio of points retained by them and the ratio of those retained by common skyline queries and ranking queries. The results of this procedure are compliant with expectations: F-Skylines are more effective than other queries and PO operator is more effective than ND operator. Moreover, another crucial result is the increase in effectiveness when the number of constraints increases. A constraint is essentially a condition thus a constraint on weights, for example, is a condition which must be fulfilled by all the functions considered ( later in this section the constraint on weights used is $W_1 \geq W_2$). The role of each constraint is reducing the space of weights, and thus the set F of functions to consider for F-dominance. It looks like a common LP problem solved geometrically.

Using a branch and bound method or a cutting plane method, constraints are added to cut the solution space and to find the desired one(or more). This is also the same method used for solving ILP problems, where the space of the solution is cut by special constraints until obtaining the required solution.

Although the increase in computational cost, this flexible skyline allows achieving the same profits as common skyline without retrieving a high size output. Just for the sake of clarity, following an example of ND and PO operators is shown. A similar example is also provided and precisely described in [3] by Ciaccia and Martinenghi. This example aims to manifest the improvements and the flexibility achieved by the previously presented operators.

Let's consider the following dataset:

| Stat | A | B | C | D | E |
|---|---|---|---|---|---|
| number of games out for injuries | 0 | 63 | 73 | 100 | 58 |
| number of red or yellow cards | 100 | 14 | 0 | 7 | 64 |

Table 10: Result of the scoring function S

Considering, that a player with a less number of games avoided due to injuries is preferred thus the preference is to have a player with the smallest number of past injuries possible. To consider this, defined the following family of scoring functions:

$$F = \{ W_1 * injuries + W_2 * redOrYellowCards : W_1 \geq W_2 \}$$

it is possible to compute the ND and the PO operator. The constraint on weights is used to model the user preference according to which the injuries weight must be higher than the red or yellow card weight. Each function belonging to F respects this constraint.

As can be seen from Figure 7(left), records A B and C are not F-dominated by any other record so they represent the output of the ND operator. It can be effectively seen, looking at figure 8, that the output of ND operator is a subset of the skyline query output. Furthermore, in Figure 7(left) it is also shown that records A and C represent also the output of the PO operator while in Figure 7 (right) B is shown to be outside PO. It can be also noticed that the output of this operator is a subset of the output of the ND operator [3]. To verify that A and C belong to the PO we need to verify that according to some weights A dominates all the other records while according to other weights, C dominates. However, B is not in PO output since it is not the top-1 result for any value of weights. Following the point B is formally proved to be dominated by A or by C while in the last inequality the point A is proved to not dominate C for every $W_1 \geq W_2$ and vice-versa, so they both are optimal solutions according to some weights.

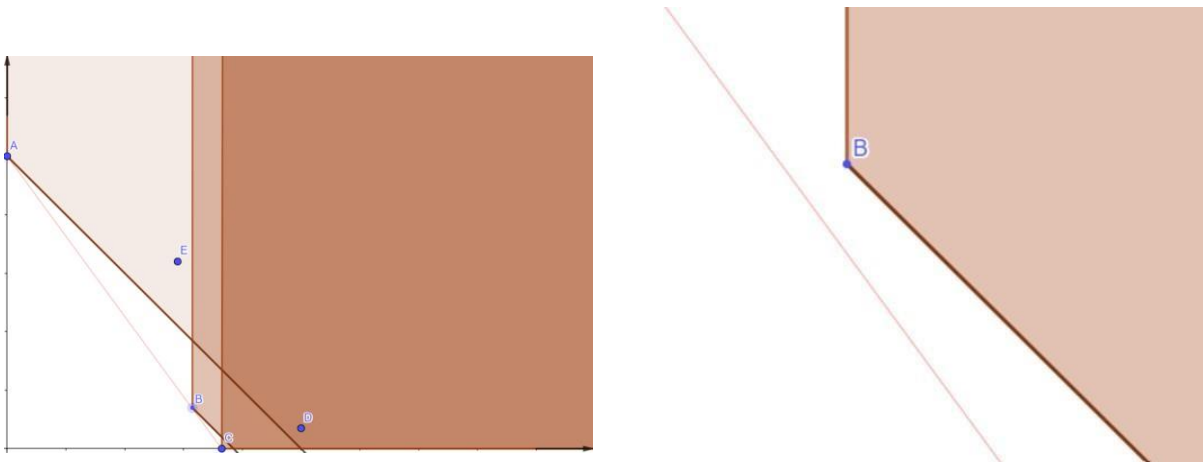

Figure 7: Dominance region (left) and enlargement of the specific point B(right)

$F(A) \leq F(B)$ then $0 * W_1 + 100 * W_2 \leq 63 * W_1 + 14 * W_2$ and so $W_1 \geq 1.365 * W_2$
$F(C) \leq F(B)$ then $73 * W_1 + 0 * W_2 \leq 63 * W_1 + 14 * W_2$ and so $W_1 \leq 1.4 * W_2$
$F(A) \leq F(C)$ then $0 * W_1 + 100 * W_2 \leq 73 * W_1 + 0 * W_2$ and so $W_1 \geq 1.369 * W_2$

So for each $W_1 \geq 1.365 * W_2$ A dominates B, while for each $W_1 \leq 1.4 W_2$ B is dominated by C so anyway B is dominated by A or C. So B is not part of PO while A and C belong to its output.

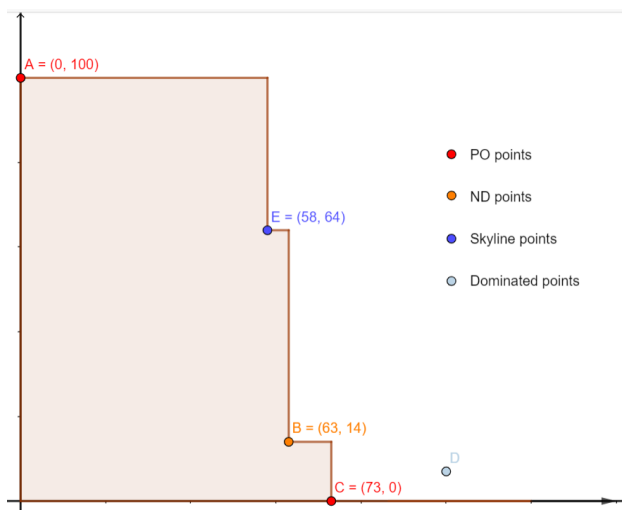

Figure 8: Skyline query of the dataset table 10

The last consideration regards the extension of the ND operator for K>1. Defined the f-skyline as the set of non dominated objects in a specific dataset D with respect to a family of function F, to specify this extension of the ND operator we require a definition of the k-Skyband and a definition of a restricted skyband with respect to F. The Skyband is the set of objects that are dominated by less than K-objects, while the restricted skyband w.r.t a family of scoring function F is a set of records which is f-dominated by less than K other records. As can be easily imagined, the restricted k-skyband is a subset of the k-skyband [5]. It can be denoted as $ND_k(D, F)$ and, when F has cardinality equal to one, thus when F is reduced to a single function, it includes all the possible top-k results for that specific F. Given these considerations, it is possible to state that, if a point P does not belong to $ND_k(D, F)$ then no scoring function in F can make P a top-k.

**Step by step. One step at a time.**

All these methods may seem to be quite uncorrelated or seem to deal with different problems, but they are often sides of the same coin. In fact, newer methods often try to expand or improve older ones. A first example is presented in the next section of the present article, where top-k queries on uncertain data (UTK) are explained. This tool tries to expand several ones and it is particularly correlated to the operators presented by Ciaccia and Martinenghi. In fact, the UTK queries use a different kind of dominance, the R-dominance, to consider a d-dimensional problem. These problem are evaluated also for $K > 1$ [18] but, for linear scoring function and for K = 1, they are equivalent to compute ND and PO operators. Further details about UTK are provided in the next section.

After the UTK, instead, there is another section dedicated to ORU and ORD. These operators, considering the results achieved by several different existing methods, try to extend them using a method based on the concept of hypersphere and even another definition of dominance, ρ-dominance [12]. For this reason, even if they are quite different methods, each one has been developed on the groundwork of the previous one.

## 4.6 Top-K queries on uncertain data

In all the previous examples, the data sets were assumed to contain available options for covering a specific need. Furthermore, the top-k queries were assumed to have a well-defined scoring function. In practice, however, this is only barely possible and there is often the need of dealing with uncertainty due to human inability or user uncertainty[18]. To do so, in the previously mentioned article the top-k queries on uncertain data (UTK) are explained and two versions of the UTK algorithm are presented.

The UTK query is strictly related to skylines but, being more precise, we should say that it is related to skyband. Later in this paper, a representation of the skyband is provided but we can consider it as a set of all those records which are dominated by less than K other records and thus it can be considered as a superset of all records that can be part of a top-k for any possible weight vector. As previously said, the idea of this tool is to unbind the top-k query from the need for a specific weight vector. The input is an entire approximated region of preference called R. This is extremely important since the input depends on this region and this region has a key role in the UTK tool. To compute the UTK output, indeed, the dominance test is not sufficient anymore. Two or more records which do not dominate a record X may still exclude it from the output if they have a better score in different parts of the region R, preventing it from entering the top-k set at any position in R. In table 11 a formalization of the problem is provided, in order to schematize all the element required to compute the UTK.

| Input: | Dataset D, integer value K > 0, preference region R |
|---|---|
| Data: | data are essentially records of a dataset. |
| | Each record Y has d values and its structure is Y = $[x_1, ..., x_d]$ |
| | These "d" attributes of Y represent a d-dimensional data domain |
| Issue on data: | There could be large datasets so a spatial index using R-Tree may be helpful |
| Weights: | $w_i$ is assumed to belong to the interval [0,1] for each $i$ belonging to [1,d] |
| | It is also assumed that $\sum_{i=1}^{d} w_i = 1$ so weights are normalized. |
| | Considering this, we can derive that $w_d = 1 - \sum_{i=1}^{d-1} w_i$ and thus |
| | the domain of w is reduced to a $(d-1)$ dimensional space called preference domain. |
| Region R: | This is the approximated region of preference given as input by the user. |

| | focus: | low-dimensional setting since, as d grows, the score of all records quickly converge and thus ranking-aware process is senseless |
|---|---|---|

Table 11: Schematisation of the elements required for UTK

Once clarified all the element involved, there are two version of the UTK which must be considered [18] and their characteristics are summarised in table 12.

| $UTK_1$ | It reports the minimal set of records that could rank among the top-k when the weight vector lies inside R. So if Y is reported then there is at least one weight vector in R for which Y belongs to the top-K set |
|---|---|
| $UTK_2$ | It reports the exact top-k set for every possible weight vector in R |

Table 12: $UTK_1$ and $UTK_2$ characteristics

In order to clarify the concept behind the UTK, following there is an example where the UTK output is computed. Let's consider the dataset in table 13 and let's compute the UTK result for k = 3.

| Stat | A | B | C | D |
|---|---|---|---|---|
| number of games out for injuries | 5 | 7 | 9 | 6 |
| number of red cards | 9 | 4 | 8 | 5 |

Table 13: Result of the scoring function S

Considering the following formula to compute the record score:
record: $Y = [x_1, ..., x_d]$

$$S(Y) = \sum_{i=1}^{d} w_i * x_i$$

and considering the following normalization on weights, such that:

$$\sum_{i=1}^{d} w_i = 1$$

since the dataset has 2 attributes, we consider the 2 dimensional problem and the previously presented formula can be seen as:

$$w_1 + w_2 = 1$$

From this, we can write the score as function of $w_1$ and substituting $w_2 = 1 - w_1$ achieving the following result:

- $x_1$ and $x_2$ are the attributes of each record
- the equation of each line representing the computation of S(Y) for each weight w1 will be:

$$S(y) = x_1 * w_1 + x_2 * (1 - w_1)$$

and so, for the dataset presented above, we have the following equations:

1. $S(A) = 5 * w_1 + 9(1 - w_1)$

2. $S(B) = 7 * w_1 + 4(1 - w_1)$

3. $S(C) = 9 * w_1 + 8(1 - w_1)$

4. $S(D) = 6 * w_1 + 5(1 - w_1)$

Records and related lines are shown in figure 9 where the horizontal axes contains all possible values of $w_1$ while the vertical axes represents the score S(Y). In this figure is also presented a possible vector w. This vector meets the lines in ascending order of score, so the top-1 tuple for the vector w presented is the last line met ( in the example, the weight vector chosen has $w_1$ = 0.27, thus $w_2$ = 0.73 and the increasing order of lines met is: B - D - A - C, so the top-1 record is C).

According to what we have said before, the top-K records for each weight are the points with at most K-1

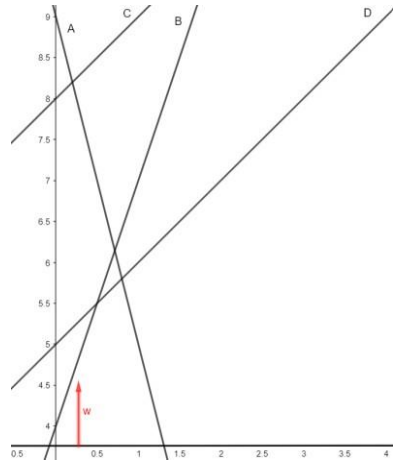

Figure 9: records A, B C and D mapped as lines

lines above ( so, the top-1 are the point with no lines above and it is called $\leq$ 1-level, the top-2 with 1 line above is called 2-level, and so on).

In this example, we are interested in finding K = 3 and, to do this, in figure 10, a region R is represented and the coloured lines represent the 1-level, the $\leq$ 2-level and the $\leq$ 3-level. The UTK is a limited version of the $\leq$ k-level and this limitation is represented by the given region R. The output of UTK for K = 3 is the portion of the $\leq$3-level inside the region R. So, in figure 10 it is the set of red points inside the region R, represented by the orange perpendicular lines.

Even if this example shows the 2-dimensional case, the same holds for any possible d, with records that are mapped into hyperplanes instead of lines. However, even if the reduction to a constrained part of the $\leq$ k-level is quite interesting, this is not a feasible way to compute UTK for each possible d since the computational effort is too high. The only possible case is for d = 2 since it degenerates into a 1-dimensional preference domain. In fact, doing the same procedure for a higher number of dimensions has a too high computational effort. For this reason, a different and more efficient algorithm is required.

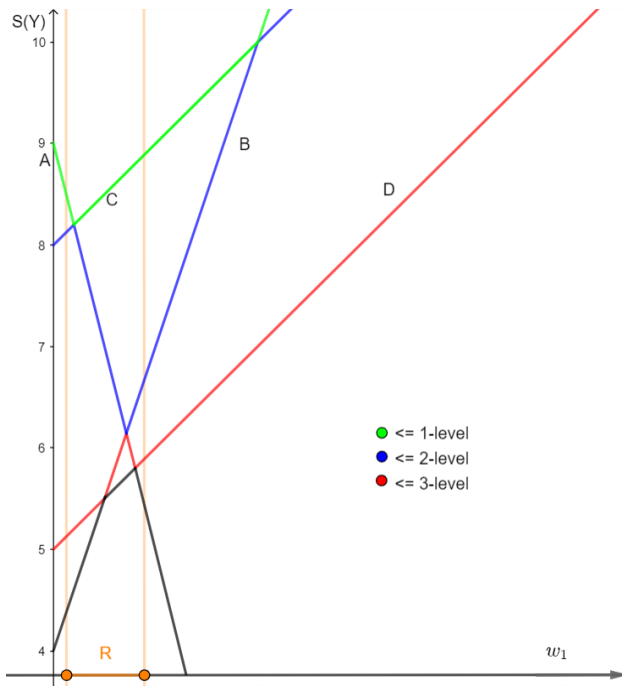

Figure 10: ≤ 1-level (green) , ≤ 2-level ( blue) and ≤ 3-level (red) representation

**Drawing a baseline**

The main idea is to construct the UTK starting from already existing methods. The first phase of the algorithm consists of a filtering step performed using K-skyband and onion layers. In order to explain what these tools are, let's consider figure 11, where a 2-skyband and a k=2 onion layers are shown to picture what they essentially represent:

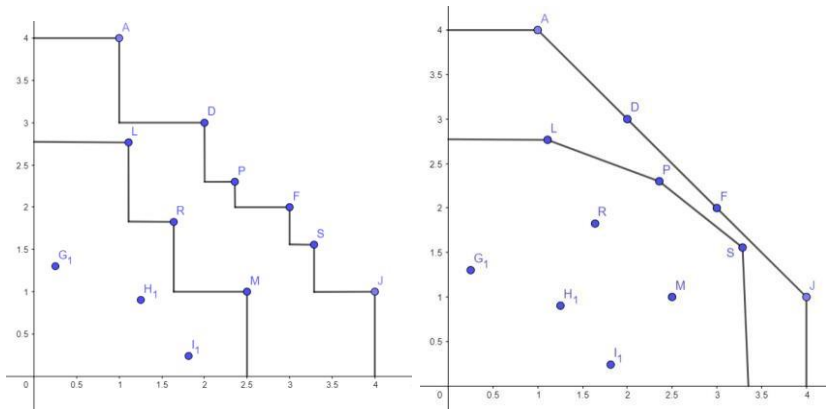

Figure 11: 2-Skyband (left) and 2 onion layers (right)

On the left, the 2-Skyband is made up of an outer line, which passes through records that are not dominated by any other record, and an inner line, which passes through records that are dominated by one record. Defined the convex hull has the smallest convex geometric structure ( the polytope) that surrounds the tuples of the dataset, in figure 11 (right) the first k=2 onion layers are shown and in its layer the records define a convex hull. Notice that implementing the onion layer from the k-skyband is better in terms of computational effort [18].

The first phase of the algorithm consists of filtering the dataset according to these two tools, which are supersets of the record that could appear in the top-k. After the first phase, each possible candidate is eval-uated to verify if it can be part of the UTK query output. To do so, the kSPR methodology [1] constrained to a specific region R is used. Since there are two different versions of UTK, we have to clarify the difference in using the above-stated methodology. In particular, for $UTK_1$ the kSPR terminates as soon as a record is found. In fact, candidate X is discarded if the output is an empty set. In $UTK_2$, instead, the kSPR is never stopped to find all the possible subregions of R in which the candidate record P could be part of the top-K result.

Considering this baseline, in [18] the RSA algorithm has been developed to evaluate UTK with less effort. The main feature of the algorithm is a different kind of dominance, the R-dominance. It would not be possible to compare tuples X and Z using the classical definition with respect to R, while it is possible using the R-dominance. Given a region R in the preference domain, a record X is said to R-dominates another record Y if the score $S(X) \geq S(Z)$ for any possible vector w belonging to the region R and there is at least a vector $w_n$ such that $S(X) > S(Y)$.

To be more precise, we can consider the equality $S(X) = S(Z)$ as an hyperplane into the preference domain and thus the inequality $S(X) \underline{\leq} S(Z)$ as an half space. Considering this, three possible situations can occur and they are shown in figure 12:

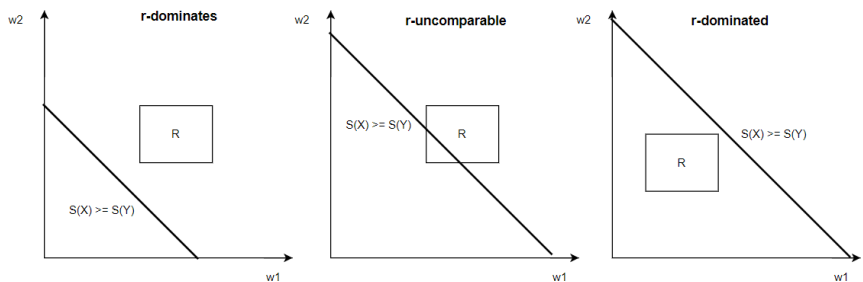

Figure 12: the 3 possible scenarios with r-dominance

The RSA algorithm is made up of a first "Filtering Phase", followed by a "Refinement Phase" and an "Optimization Phase"[18]. In each phase, the concept of R-dominance is used to compare tuples and find the dominated and non dominated ones.

So, while the ND and PO operators previously seen allowed to evaluate the case of K = 1, this method extends that concept to a more general situation of K > 1 and it has several modern applications like data cleaning or data integration [22]. Moreover, a detailed analysis of the score distribution is performed in the previously mentioned article, enriched by several examples aimed to avoid ambiguity.

## 4.7 ORU and ORD operators

Performing an analysis of skyline queries and ranking queries, it is possible to identify three hard requirements for the solution of the multi-criteria decision problem[12]. One of these has been already anticipated before, while the others are here are reported in order to lay the groundwork for understanding the subsequent operators shown:

1. The second one concerns the possibility of controlling the output size
2. The third one concerns the flexibility in preference specification.

Combining them allows the creation of a flexible skyline query that faces disadvantages of the all previously described queries. The tool presented to fulfil the requirements described are the ORU and ORD operators [12] .

### 4.7.1 ORD operator

The first operator, ORD, reports the records that are $\rho$-dominated by fewer than K others, for the minimum $\rho$ that produces exactly m records as output. In order to work properly, it requires a seed vector w and the desired output size.

For sake of clarity, it is needed to define the meaning of $\rho$-domination. As declared in the aforementioned article, let w be a best-effort estimate of the user's preference vector and let consider the preference vectors v within distance $\rho$ from w, i.e., where $|v - w| \leq \rho$. Considering this, a record x $\rho$-dominates another record y if it scores at least as high as y for every such vector v, and strictly higher for at least one of them. So, while in the previous section the concept of F-dominance was centred on a simple dominance of tuples, in the $\rho$-domination a tuple dominates another one if and only if the dominating is always at least as good as the dominated one and is better than the dominated one for at least a parameter. Inside the article is also shown a framework that can define the right $\rho$ according to the desired size of the output m. The method suggested consists on using an incremental $\rho$-skyline algorithm to compute a $\rho$. In particular, the algorithm is incrementally repeated until it has m records and the $\rho$ will be the final radius reported.

To permit a first evaluation of the inflection radius, whose further details are better explained in [12], it is possible to consider these records:

| Stat | AAA | BBB | CCC | DDD | EEE | FFF | GGG |
|---|---|---|---|---|---|---|---|
| number of games out for injuries | 27 | 48 | 53 | 56 | 54 | 57 | 58 |
| number of red cards | 6 | 3 | 5 | 2 | 4 | 1 | 0 |

Table 14: Example of dataset

Using this dataset, it is computed the skyline query to point out which are the non-dominated tuples:

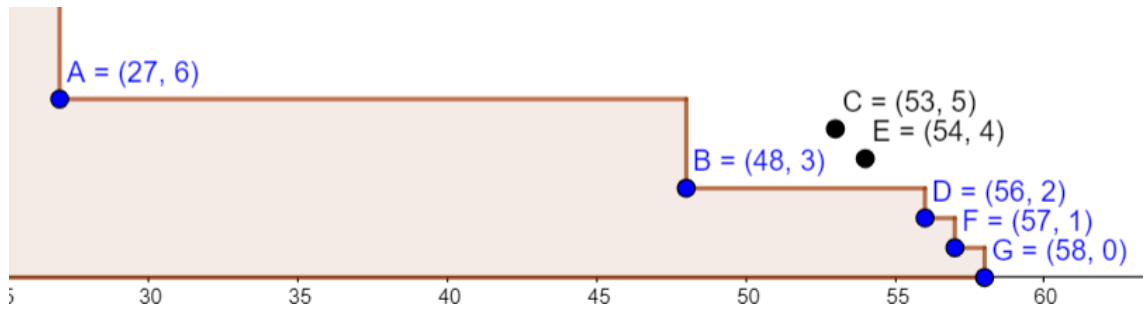

Figure 13: Skyline Query

From the dataset, it is evident that the non-dominated elements are: AAA,BBB,DDD,FFF,GGG According to a specific weight vector, e.g. W = [1,2], the first step consist of computing the score for each one of the tuples in the dataset:

| Stat | AAA | BBB | CCC | DDD | EEE | FFF | GGG |
|---|---|---|---|---|---|---|---|
| number of games out for injuries | 27 | 48 | 53 | 56 | 54 | 57 | 58 |
| number of red cards | 6 | 3 | 5 | 2 | 4 | 1 | 0 |
| Score | 39 | 54 | 63 | 60 | 62 | 59 | 58 |

Table 15: Example of dataset after calculating the score

Now, to proceed with the example, there have to be considered the following three categories while for the computing of the inflection radius, the record BBB is considered.

1. First category: it contains all the record which scores less than the non-dominating ones since these records will never dominate them for any value of $\rho$

2. Second category: it contains all the record that dominates the considered one with the traditional skyline query. In fact, these records dominate the considered one for each value of $\rho$.

3. Third category: It contains all the remaining records, so all the records that do not dominate the considered one but which perform better according to the considered weight vector.

As can be easily imagined, the focus of the algorithm regards the third category since they are proved to dominate the examined one for a finite value of $\rho$.

The third category of records contains DDD, FFF and GGG which do not dominate the investigated one, BBB, but which perform better according to the weight vector chosen.

Starting from the previously used weight vector, w = [1,2], the first value of $\rho$ studied is $\rho$ = 1. According to it, the preference vector v can be computed by adding in each direction the $\rho$ value to the weight vector. In particular:

(1,2) + (-1,0)= (0,2)
(1,2) + (1, 0)= (2,2)
(1,2) + (0,1) = (1,3)
(1,2) + (0,-1)= (1,1)

So the v vector is made up of the following points:
$$v = [(0, 2), (2, 2), (1, 3), (1, 1)]$$
Fixing K = 3 and considering the first element of the v vector ( the element (0,2) ), in the following tables the values of the new score are stored:

| Stat | AAA | BBB | CCC | DDD | EEE | FFF | GGG |
|---|---|---|---|---|---|---|---|
| number of games out for injuries | 27 | 48 | 53 | 56 | 54 | 57 | 58 |
| number of red cards | 6 | 3 | 5 | 2 | 4 | 1 | 0 |
| Score | 12 | 6 | 10 | 4 | 8 | 2 | 0 |

Table 16: Example of dataset after calculating the score according to v1 = (0,2)

This time, only FFF has the same score as BBB. So, since there are less than K values which $\rho$-dominate the considered one, the algorithm is over and the inflection radius, which is the value of $\rho$ when the algorithm is completed, is equal to 1.

Actually, the best performing algorithm can evaluate the scoring function and the inflection radius on the fly and it iterates several times the computation for different weight vectors [12].

### 4.7.2 ORU operator

The second operator, ORU, follows more closely the ranking by utility paradigm than ORD. This operator reports the tuples which belong to the top-k results obtained using at least one preference vector within the distance $\rho$ from w, for the smallest $\rho$ value that has a size output of exactly m.

One of the main strengths of both the operators is the absence of precomputation beyond a general-purpose index on the dataset. This is relevant since the updates on D affect only the index and the integration of common predicates into the shown frameworks is possible[12].

To estimate the ORU operator, it is required the computation of the convex hull of the dataset. Starting from it, considering a specific weight vector W, it allows to compute the $\rho$-skyline query more geometrically. For the sake of clarity, in the figure 14 an example of convex hull is shown in order to explain how it is used by the operator.

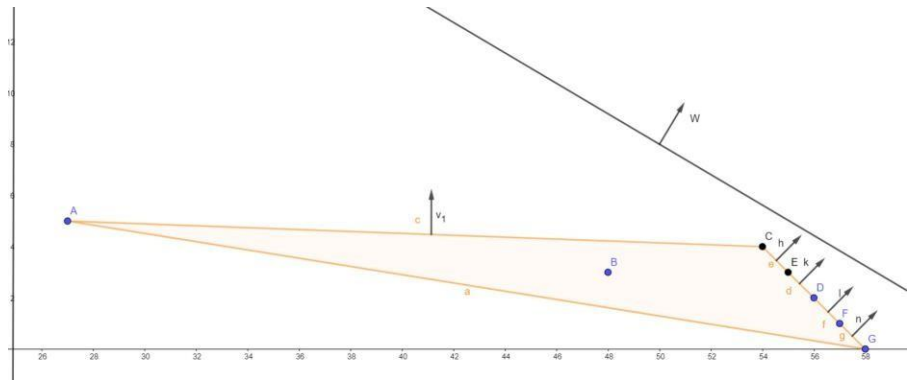

Figure 14: Convex Hull of the dataset

The upper hull of the convex hull is composed of the facets AC, CE, ED, EF, EG ( orange labels in the figure). Moving the hyper-plane, which is orthogonal to W, from top to the origin, it is quite simple to see that the records are reached in the following order:
C - E - D - F - G - A
Given this order, it is possible to map each facet reached on the figure 15: In this figure, the difference between the orientation of the line orthogonal to W and the orientation of the facet is represented by the distance from the vector w, which is pictured by the point w. On this figure, it is possible to choose a value of $\rho$ and, starting from the top record, it is evidenced which records are contained inside the region described by $\rho$. Note that, since some facets have the same orientation, some vectors are parallel to each other and so some points are overlapped.

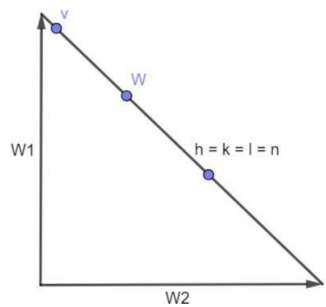

Figure 15: Preference region according to the vector w

### 4.7.3 ORU and ORD comparison

To sum up, the introduced algorithms are proved to be practical and scalable and, even if it is shown for a low number of dimensions, it lays the groundwork for studying also a highly deflected or sparse set of records, where multi-objective querying should be allowed for a greater number of dimensions.

# 5 Comparison and Conclusions

As a result of the previously described method, it is now possible to precisely define the difference between skyline queries and flexible skyline queries and it is possible to highlight the advantages provided by the second one.

Before putting a spotlight on such difference, it is imperative to remark on the concept of domination among the previously seen methods to emphasize the difference among them. In table 17, the distinct definitions are reported together in order to simplify the comparison among them.

| Name | Definition |
|---|---|
| Dominance | A record X is said to dominate a record Y if it is better than Y according to certain weights |
| R-Dominance | Given a specific region in the preference domain called R, a record X is said to r-dominate Y if it is at least as preferable as Y for each weight vector belonging to R and there exists at least a weight vector according to which X is better than Y |
| Trade-off dominance | A tuple X is said to dominate Y with respect to the trade-off dominance if it dominates Y according to a specific trade-off set T. |
| $\epsilon$-Dominance | A tuple X is said to be dominated by Y if it belongs to the domination region of Y increased (or reduced) of a certain area $\epsilon$ |
| F-Dominance | Defined a family F of scoring function, a tuple X is said to F-dominate another record Y when X is always better than Y according to every function in the family F |
| $\rho$-Dominance | A tuple X is said to $\rho$-dominate another tuple Y if it is always at least as good as Y and it is better than Y for at least a parameter |

Table 17: Comparative table of different dominance definitions

Using this table to lay the groundwork for the comparison, it is very clear that each definition is an improvement or a modification of the previous one. For the sake of clarity, let's have a short comparison of all these different definitions of dominance.

The first one is the classic definition, which represents the key to the classic skyline framework. It simply consists of evaluating two records given specific weights. Starting from this definition, the others have been developed to fit different requirements or highlight different properties. The easiest and most intuitive is the $\epsilon$- dominance, which is essentially an extension or a reduction of the dominance region, depending on the value given to $\epsilon$. In particular, given a negative value of $\epsilon$ it is possible to enlarge the dominance region (this is the only method which allows an enlargement). Other definitions like the F-dominance, the R-dominance and the $\rho$-dominance are more sophisticated. In particular, the F-dominance takes into account an entire set of possible scoring functions to compute ND and PO operators and its strength is the ability to consider user preferences. Since F-skylines has been generalized by the uncertain top-k queries, it is reasonable that R-dominance and F-dominance are tightly linked. In fact, given a set of linear scoring functions, the ND re- sult for K = 1 is equivalent to computing part of the UTK result, which allows the evaluation of the problem also for K > 1. The R dominance consists of evaluating several weight vectors belonging to a specific region R. The last kind of dominance has some similarities with F-dominance since they both consider a family of the preference input, but the main peculiarity is considering a specific restriction $m$ over the output size. Moreover, considering the preference domain, $\rho$ of the $\rho$-dominance can be represented as the radius of a hypersphere centred in the specific weight vector considered.

As for these definitions, also the methods are different but not uncorrelated to each other and they draw a sufficiently marked line between non-flexible and flexible skyline.

As suggested by the name, flexible skyline queries are less strict than the classic skyline query and they allow facing one or more downsides of the classic tool.

Following, taking into account all the considerations made about the methods presented in the previous sections of the present article, a comparison among f-skyline, skyline queries and ranking queries in terms of computational effort is performed.

On the one hand, the computational effort required by an f-skyline and skyline is essentially the same so there is no reason to prefer a skyline to an f-skyline.

On the other hand, the computational effort required by f-skyline is bigger than the one required by ranking queries, so the flexibility has enough significant drawbacks. However, the advantages achieved using f-skyline are important and their use is justified also despite the computational effort.

The last two operators are ORD and ORU which are presented in [12]. They were thought to bring together the strengths of the two main previously seen paradigms, the dominance-based and ranking by a utility function. The goal of this union was to avoid the drawbacks of both paradigms. As a result, all the issues of ranking and skyline queries are mostly overcome.

In conclusion, the new operators, ORU and ORD, perform as good as expected and their supremacy in terms of computational effort and quality of the results is proved. Moreover, it is remarkable that these operators sometimes include records that are missed by common skyline queries. As a result of all these considerations, the improvement achieved with f-skyline queries should now be clear, having also considered how the issues have been mostly solved without losing the benefits of the classical tools.